\documentclass[intlimits,twoside,a4paper]{article}

\usepackage{amsmath,amssymb}
\usepackage{graphicx,xcolor}

\usepackage[T2A]{fontenc}
\usepackage[cp1251]{inputenc}

\usepackage[normal]{subfigure}		

\usepackage[eqsecnum]{cmpj2}

\issue{2014}{17}{3}{33603}
\doinumber{10.5488/CMP.17.33603}

\newcommand{\sysb}{\left\{\begin{array}}        

\newcommand{\syse}{\end{array}\right.}			

\newcommand{\reff}[1]{(\ref{#1})}				

\newcommand{\suml}[2]{\sum_{#1}^{#2}}			

\newcommand{\dd}[2]{\frac{\rd^{#1}#2}{\left( 2\pi \right)^{#1 }}}   

\newcommand{\order}[1]{O\left( #1 \right)}		

\newcommand{\wt}{\widetilde}					

\newcommand{\tx}{\tilde{x}}						
\newcommand{\ty}{\tilde{y}}

\newcommand{\lt}{\left(}						
\newcommand{\rt}{\right)}						
\newcommand{\lqq}{\left[}						
\newcommand{\rqq}{\right]}						
\newcommand{\lan}{\left\langle}					
\newcommand{\ran}{\right\rangle}				
\newcommand{\abs}[1]{\left| #1 \right|}			
\newcommand{\eval}[1]{\left.\right|_{ #1 }}		
\newcommand{\av}[1]{\lan #1 \ran}				

\newcommand{\cosha}[1]{\cosh \left(  #1 \right)}		
\newcommand{\sinha}[1]{\sinh \left(  #1 \right)}		

\newcommand{\lna}[1]{\ln \lt #1 \rt}					
\newcommand{\BK}[2]{K_{#1} \lt #2 \rt}					

\newcommand{\EGamma}[1]{\Gamma \lt #1 \rt}				

\newcommand{\rme}[1]{{\re}^{#1}}				

\newcommand{\change}[1]{\textcolor{black}{#1}}

\title[Critical relaxation and the combined effects of spatial and temporal boundaries]%
{Critical relaxation and the combined effects \\ of spatial and temporal boundaries}

\author[M. Marcuzzi, A. Gambassi]{M. Marcuzzi\refaddr{label1,label2},
A. Gambassi\refaddr{label1}}
\addresses{
\addr{label1} SISSA --- International School for Advanced Studies and INFN, via Bonomea 265, 34136 Trieste, Italy
\addr{label2} School of Physics and Astronomy, University of Nottingham, University Park, Nottingham, NG7 2RD, UK
}

\date{Received May 6, 2014, in final form June 12, 2014}

\begin{document}

\maketitle

\begin{abstract}
\change{We revisit here the problem of the collective non-equilibrium dynamics of a classical statistical system at a critical point and in the presence of surfaces.
The effects of breaking separately space- and time-translational invariance are
well understood, hence we focus here on the emergence of a non-trivial interplay between them. For this purpose, we consider a
\change{semi-infinite} model with $O(n)$-symmetry and purely dissipative dynamics which is prepared in a disordered state and then suddenly quenched to its critical temperature. We 
determine the short-distance behaviour of its response function
within a 
perturbative approach
which does not rely 
on any a priori assumption on the scaling form of this quantity.}
\pacs 64.60.De, 64.60.Ht, 68.35.Rh
\end{abstract}


\section{Introduction}

It is a trivial observation that any physical system has actually a finite extent;
as a consequence, descriptions which assume translational invariance can at most capture
its
bulk features, with surface effects representing subleading corrections which decay upon moving away from the boundaries.
This decay is controlled by the presence of an inherent length scale, which sets the ``range'' of these surface effects. Rather generically,
such a scale corresponds to the correlation length $\xi$ of the system, which encodes the separation beyond which
different regions of the extended system are no longer statistically correlated.
Accordingly, one can identify three distinct instances:
\begin{itemize}
	\item[(i)] $\xi$
	is comparable with the typical linear extent $L$ of the sample and the behaviour of the system is strongly affected by its finiteness, as every point
	effectively feels the presence of the boundaries. In this case finite-size effects emerge \cite{Barber-rev} and the thermodynamic quantities
	explicitly depend on $L$.
	\item[(ii)] $\xi$ is significantly smaller than $L$ but one focuses on the behaviour at spatial points located at a distance $d \gg \xi$ from the
	boundaries, in such a way that the effects of the boundaries can be neglected and the system can be modelled as being infinitely extended, with
	translationally invariant properties.
	\item[(iii)] $\xi$ is much smaller than $L$ but one considers the behaviour of the system within a
	distance $d \ll \xi$ from the boundaries. If $\xi$ is smaller than the curvature of the latter and there are no wedges or tips,
	a suitable description is provided by semi-infinite models with a flat boundary and a lingering (approximate) translational invariance in all spatial
	directions parallel to the surface.
\end{itemize}
In what follows, we focus on case (iii).
The differences between the behaviour close to the boundaries and the one far from them, i.e., in the bulk, formally arise due to
the explicit breaking of the translational invariance along the direction orthogonal to the surface, which allows an extended freedom in the system:
for example, a two-point correlation function $C(x,y)$ is no longer constrained to be a function of the distance $\abs{x-y}$.

A faithful description of these \change{systems} generically requires accounting for all the microscopic features which characterize both the bulk and the surface; in turn, this implies that the corresponding behaviour is highly system-specific. However, when the correlation length $\xi$ becomes large with respect to microscopic scales, collective behaviours emerge which are not determined by the underlying microscopic structure, but by those coarse-grained properties which do not really depend on the considered scale, such as symmetries, range of interactions and the (effective) spatial dimensionality.
These circumstances, which represent a hallmark of systems undergoing a continuous phase transition, lead to \emph{universality}. In other words, several relevant quantities can be identified which, due to the very fact that the microscopic details become inconsequential for their determination, take
the same values in many different systems, which in turn share the same gross features and constitute as a whole the so-called \emph{universality class} of the transition.
Consequently, it is sufficient to study just one representative system in order to gain information on the whole class it belongs to.
Continuous phase transitions are typically associated with the \emph{spontaneous breaking} of some underlying symmetry \cite{Zinn-Justin},
which is highlighted by the behaviour of the so-called order parameter $\varphi$ (e.g., the local magnetisation for an Ising ferromagnet) upon crossing the critical point.
The emergence of universality is currently understood within the framework of the renormalization group (RG) \cite{Wilson, Ma, Zinn-Justin}. Its transformations act by enlarging the length scale at which a system is described, progressively blurring details at shorter scales.
Under the assumption that $\xi$ represents the only inherent non-microscopic length scale, one must conclude that its divergence deprives the system of any typical scale and, therefore, its physical behaviour should become self-similar under scale dilatations, and the RG transformations reach a fixed point. The different physical systems which end up falling into it constitute a certain universality class. As a matter of fact, as long as the interest lies in the study and determination of universal quantities, the resulting mesoscopic description can be formulated in terms of fields on a space-time continuum, which makes it possible to use standard field-theoretical methods in order to calculate many relevant quantities \cite{Zinn-Justin}.

The discussion below complements the results presented in reference \cite{M-class} by providing in detail the analytical derivation of the first
corrections to the linear response function of a system bounded by a flat surface and subject to a temperature quench.
In particular, in section \ref{sec:crit} we briefly discuss the main features emerging in critical systems with spatial and temporal boundaries, while in section \ref{sec:model} we set up the description of the aforementioned model in terms of a suitable (field) theory on the continuum; by using standard field-theoretical methods and a RG-improved perturbation theory we explicitly calculate the relevant universal quantities and show the emergence of an unexpected edge behavior. Finally, in section \ref{sec:concl} we draw our conclusions.

\section{
\change{Equilibrium transitions at surfaces and non-equilibrium critical dynamics after a quench}}
\label{sec:crit}

\subsection{\change{Spatial boundaries: Equilibrium critical behaviour at surfaces}}
\label{ssec:surf}

Being originally devised for describing the behaviour of unbounded and uniform systems,
which provide a good approximation of regime (ii) above, the RG has been subsequently
generalized in order to describe finite-size effects (regime (i)) \cite{Barber-rev} and the presence of boundaries \cite{Binder-rev, Diehl-rev, Diehl2, Pleim-rev} (regime (iii)). In the latter cases, the breaking of translational invariance plays a fundamental role, leading to the appearance of novel scaling behaviours which can be observed, for example, in the correlation functions of the order parameter $\varphi$ upon approaching the boundaries \cite{Binder-rev, Diehl-rev, Diehl2, Pleim-rev, CritDyn-Gambassi} and are characterized by new universal exponents which cannot be inferred from the bulk ones.
Within the RG approach, the additional parameters which describe the gross features of a
boundary (e.g., a different interaction strength at the surface with respect to that in the bulk)
typically \change{give rise to} a splitting of the original bulk universality class into
a set of surface subclasses characterized by these \change{novel} (surface) exponents.
A number of analytical \cite{Special,Ordinary,Diehl2}, numerical \cite{Hasen_sp,Num4,Num5}
and experimental (see, e.g., reference \cite{ExpRef}) studies have been carried out to investigate
semi-infinite and film geometries, whereas wedges, edges, and the associated
 critical Casimir forces \cite{Cardy-edge,Pleimling3,Dietrich1},
as well as curved and irregular surfaces \cite{Diehl-rev,Pleim-rev, Hanke1} have been studied to a lesser extent.

In order to \change{exemplify} some of the features mentioned above, we focus now on the Ising universality class, which is effectively described by an effective Ginzburg-Landau free-energy
\begin{equation}
\mathcal{F} [\varphi] = \int \rd^d x \lqq   \frac{1}{2} \lt \vec{\nabla} \varphi \rt^2 + \frac{r}{2} \varphi^2 + \frac{g}{4!} \lt  \varphi^2  \rt^2 \rqq ,
\label{eq:Stree}
\end{equation}
where \change{$r \propto T - T_\textrm{c}$ controls the distance (in temperature) from the critical point} and $g>0$ sets the strength of the interaction.
In the presence of a surface, the spatial integration above is restricted to \change{the half-space} $x_\perp >0$, $x_\perp$ being the coordinate in the direction orthogonal to the surface. Furthermore, a term $\mathcal{F}_1$ which accounts for the surface properties has to be added to equation
\reff{eq:Stree}.
In view of the eventual application of RG, one can conveniently focus on the most relevant
surface term which is allowed by dimensional analysis: it turns out to have the form
(assuming that the bulk symmetry $\varphi \leftrightarrow -\varphi$
 is not explicitly broken at the boundary)
\begin{equation}
	\change{\mathcal{F}_1 = \int\! \rd^{d-1} x \, \frac{c_0}{2} \, \varphi^2(x_\perp = 0),}
	\label{eq:SS}
\end{equation}
\change{where the integration only runs on the coordinates parallel to the surface,} and effectively encodes the boundary condition \cite{Diehl-rev}
\begin{equation}
	\partial_{x_\perp} \varphi \eval{x_\perp = 0} = c_0\,  \varphi(x_\perp =0).
	\label{eq:BC}
\end{equation}
The \emph{surface enhancement} $c_0$ accounts for the differences in the interaction strength between the bulk and the surface.
\begin{figure}[!b]
\centerline{
\includegraphics[width=0.6\columnwidth]{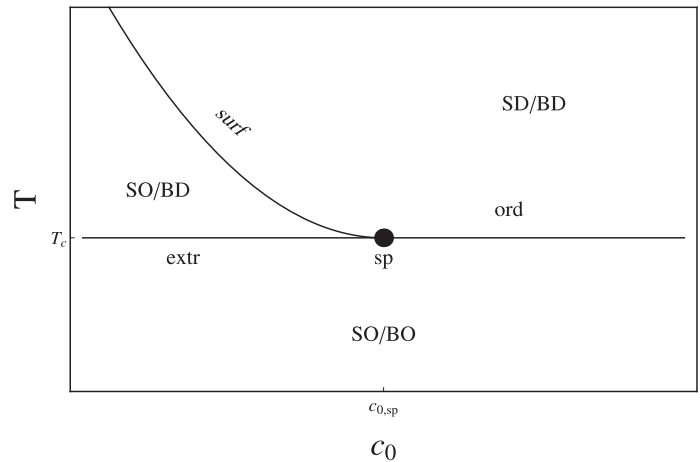}
}
\caption{Qualitative temperature ($T$)~--- surface enhancement ($c_0$) phase diagram
of the Ising model. Here S stands for ``surface'', B for ``bulk'', O for ``ordered''
and D for ``disordered''. The point $c_{0,\textrm{sp}}$ denotes the special
transition and $T_\textrm{c}$ the critical temperature. Departing from this point one
identifies three lines corresponding to the ordinary (ord), extraordinary (extr) and surface (surf) transitions.}
		\label{fig:PD}
		\end{figure}
Its presence modifies the characteristic phase diagram of the Ising model, as shown in figure~\ref{fig:PD}.
In addition to the usual ferromagnetic (SO/BO in the figure) and paramagnetic (SD/BD) phases, one may find
a third one (SO/BD) in which the surface (S) is ordered (O) whereas the bulk (B) is disordered (D).
Accordingly, the number of different transitions (and, therefore, universality classes)
increases to four and they are referred to as \emph{ordinary} (the surface orders with the bulk),
\emph{surface} (only the surface becomes ferromagnetic), \emph{extraordinary}
(the bulk orders in the presence of an already magnetized surface) and \emph{special}
(\change{the point at which the critical lines coalesce}) \cite{Binder-rev, Diehl-rev}.
Note that every transition except for the ordinary one inherently requires the possibility
for the surface to order independently of the bulk, i.e., that its dimensionality $d-1$
is strictly larger than the lower critical dimension $d_\textrm{lc}=1$ of this system.
Thus, semi-infinite ferromagnets in $d=2$ can only undergo the ordinary transition.
\change{We recall that for continuous symmetries (e.g., $O(n)$ with $n \geqslant 2$)
one has instead $d_\textrm{lc} \geqslant 2$ \cite{Zinn-Justin}}.
The ordinary and special transitions take place with a vanishing order parameter both in the bulk and at the surface;
thanks to this, they admit a unified
description which differs only by the effective boundary conditions (see equation \reff{eq:BC})
cast onto the order parameter, which within the Gaussian
approximation are of Dirichlet (i.e., $c_{0,\textrm{ord}} = +\infty$) and Neumann ($c_{0,\textrm{sp}} = 0$) type, respectively.
As stated above, the magnetization $m_\textrm{s}$ at the surface [with $m(x) = \langle \varphi(x) \rangle$] and the one $m_\textrm{b}$
in the bulk show different \change{behaviours when varying the temperature $T$ in the critical regime}, i.e.,
\begin{equation}
	m_\textrm{b} \propto \abs{T - T_\textrm{c}}^\beta, \quad \quad m_\textrm{s} \propto \abs{T - T_\textrm{c}}^{\beta_1},
	\label{eq:diffsc}
\end{equation}
which require the introduction of a new critical exponent $\beta_1$; this exponent is also reflected in the generic dependence of $m$ on $x_\perp$ upon approaching the surface, which is determined by the so-called \emph{short-distance expansion} (SDE) \cite{Diehl-rev}
\begin{equation}
	m (x_\perp \to 0) \sim x_\perp^{\frac{\beta_1 - \beta}{\nu}} m_\textrm{s} \,.
	\label{eq:SDE}
\end{equation}
Thus, one can characterize the critical surface behaviour by means of the algebraic dependence
of certain quantities on the distance from the surface.

\subsection{
\change{Temporal boundaries: short-time critical dynamics after a temperature quench}}
\label{ssec:tempora}
Rather surprisingly, the formalism described
 above can be applied within \change{a dynamic framework as well, after performing a sharp variation of one of the thermodynamic control parameters of the system (e.g., the temperature) \cite{Janssen}}.
\change{Intuitively}, the instant $t_0$ at which such a quench is performed separates the equilibrium regime ($t < t_0$) from the non-equilibrium one ($t> t_0$) and, therefore, acts qualitatively exactly in the same way as a spatial surface which stands between the outside and the inside of a system.
In this case, the distance $x_\perp$ from the surface is given by the time $t-t_0$ elapsed from the quench.

\change{The emerging scaling properties of the dynamics of a statistical system near criticality are more easily discussed in terms of the evolution of the associated order parameter field $\varphi$ on the continuum. For a broad class of systems (the so-called \emph{model A} universality class in the notion of reference \cite{HH}, see further below), the features of such an evolution can be effectively captured via a Langevin equation of the form
\begin{equation}
			\frac{\partial\varphi}{\partial t} = - \Omega \frac{\delta \mathcal{F}}{\delta\varphi} + \eta ,
		\label{eq:Langevin}
\end{equation}
where $\mathcal{F}$ is the
effective free-energy, $\Omega$ is a kinetic coefficient, while $\eta$ is a Gaussian white noise with
\begin{equation}
	\av{\eta(t)}=0   \qquad \text{and} \qquad   \av{\eta(t) \eta(s)} = 2 \Omega k_\textrm{B} T \delta (t-s),
\end{equation}
which accounts for the fluctuations due to a thermal bath at temperature $T$ and represents an external source of dissipation (for simplicity, units are chosen so that $k_\textrm{B} T =1$).
This kind of equations can be mapped onto a field-theoretical description \cite{MSR,MSR+J,MSR+D} via the introduction of the so-called \emph{response field} $\widetilde{\varphi}$. The corresponding action reads
\begin{equation}
	S[\varphi, \widetilde{\varphi},  \eta] = \int \rd^d x\, \rd t \, \widetilde{\varphi} \lt \partial_t\varphi + \Omega \frac{\delta \hat{\mathcal{F}}}{\delta\varphi}  - \eta \rt.
	\label{eq:act1}
\end{equation}
The newly-introduced variable $\widetilde{\varphi}$ actually encodes the response properties of the system:
indeed, $\av{O \widetilde{\varphi} (s)}$ is proportional to the response of the observable $\av{O} \equiv \int [\rd \varphi \, \rd \widetilde{\varphi} \, \rd \eta]\, O \,\re^{-S[\varphi,\widetilde{\varphi}, \eta]}$ to an external perturbation applied at time $s$ which couples linearly to $\varphi$. In fact, 
introducing a source $\mathcal{F} \to \mathcal{F} - \int \rd^d x \, h \varphi$ yields
\begin{equation}
	\frac{\delta \av{O}}{\delta h(s)}\Big|_{h \equiv 0} =\Omega \av{O \widetilde{\varphi}(s)}.
\end{equation}
Analogously to the spatial case, the quench limits the integration to times $t\geqslant t_0$. Moreover, the determination of the dynamics needs an initial condition $\varphi_0$ for the field $\varphi$ to be specified, for example via its probability distribution $\mathcal{P} (\varphi_0)$, which can be conveniently written in the exponential form 
$\mathcal{P} (\varphi_0) = \rme{-S_0 \lqq \varphi_0 \rqq}$.
The associated boundary action $S_0$ is integrated only over spatial coordinates and is thus akin to $\mathcal{F}_1$ in the spatial case illustrated in section~\ref{ssec:surf}. By an RG argument, one can analogously account only for the most relevant terms in $S_0$; considering again the Ising universality class as an example, one would then have
\begin{equation}
	S_0 \lqq \varphi_0 \rqq = \int \rd^d x\, \lt   \tau_0 \varphi_0^2 + g_0 \varphi_0^4 \rt.
\end{equation}
With the additional assumption that the initial state is very far from criticality, one can also neglect the second addend and thus obtain a Gaussian initial condition with variance $\tau_0^{-1}$.
Although the structure of $S_0$ looks the same as the one of ${\mathcal F}_1$ in equation~\reff{eq:SS}, one has to take into account the fact that $\mathcal{P} (\varphi_0)$ is a probability and
thus $\tau_0$ can be neither vanishing nor negative.
Hence, without explicitly breaking the Ising symmetry $\varphi_0 \leftrightarrow -\varphi_0$ of 
the action, the only fixed (i.e., critical) point one can identify is the equivalent of the ordinary one with  
$\tau_0 \to +\infty$ \cite{Janssen}, corresponding to a state with vanishing correlations, typically related to a very high temperature.}

\change{Depending on the gross features of the dynamics, such as conservation laws etc., a splitting of the equilibrium universality classes is found \cite{HH}, associated}
to the appearance of new universal quantities, such as the dynamical exponent $z$ which encodes the difference in scaling dimension between space and time coordinates \cite{HH} and thus describes how the typical linear relaxation time $t_R\sim \xi^z$ grows upon approaching the critical point. For example, a classical Ising model on a lattice which evolves in time via thermally-activated independent flips of the spin (Glauber dynamics) belongs to the so-called  \emph{model A} universality class (in the notion of reference \cite{HH}) and is characterized by having, within the Gaussian approximation, $z=2$.
Conversely, if the evolution conserves the total magnetization (Kawasaki dynamics), i.e., it makes domains diffuse or split, or combine, the universality class changes to \emph{model B}, with Gaussian dynamical exponent $z=4$, whilst all the universal features of time-independent quantities remain the same. The universal behaviour emerging at a "temporal" boundary takes the form of an
\emph{initial slip} and emerges as discussed above for spatial boundaries, i.e., because of the different scaling behaviour of observables in the initial and final stages of the non-equilibrium dynamics. In fact, analogously to equation \reff{eq:SDE}, this behaviour
can be extracted from the short-distance expansion observed for $t \to t_0$ \cite{Janssen}.

\section{
\change{Quenching a dynamical model with a surface: emergence of edge  effects}}
\label{sec:model}

\change{We consider here a purely dissipative dynamics in the presence of a spatial surface, which can be described by model A reported in equation~\reff{eq:Langevin} with the effective Ginzburg-Landau free energy $\mathcal F$ replaced by $\hat{\mathcal{F}} = \mathcal{F} + \mathcal{F}_1$,}
which includes the \change{semi-infinite bulk} and the surface contributions $\mathcal{F} $ (equation \reff{eq:Stree}) and $ \mathcal{F}_1$ (equation \reff{eq:SS}), respectively.
For the purposes of the present analysis, the kinetic coefficient $\Omega$ can be generically set to 1
by rescaling time and noise as $t \to t/\Omega$ and $\eta \to \Omega \eta$, respectively.
In order to introduce the temporal boundary, we consider the system to be prepared in a completely disordered state for $t<t_0=0$, which corresponds to vanishing $\varphi$ and, in turn, to Dirichlet boundary conditions for the dynamics \cite{Janssen,CritDyn-Gambassi}.
For the spatial boundary, instead, we will consider both the ordinary and special points.

The separate effects on the system of  either spatial or temporal boundary
\cite{Diehl-rev, Special, Ordinary, Diehl2, Janssen, Pleimling, EqSurface, NonEqFinite},
including the case in which an initial non-vanishing magnetisation is present \cite{DietrichNot,Dietrich2},
are  well-understood; on the other hand,
their interplay has been less extensively studied \cite{M-class, Ritschel, Sengupta, CritDyn-Gambassi}. Moreover, in reference \cite{Ritschel}, a power-counting argument has been used for arguing that no new algebraic behaviours arise which are specific to the intersection of the two boundaries (hereafter referred to as the \emph{edge}) and that all the observed effects are, therefore, a combination of those independently generated by the surface and the quench;
almost all the subsequent studies formulated scaling ansatzes based on this assumption.
The analysis of reference \cite{M-class}, \change{instead}, is not based on it but proceeds to a direct calculation of the effects of the edge. As we detail below, this leads to the emergence of new (field-theoretical) divergences which are sharply localised at the edge and, therefore, highlight non-trivial modifications to the scaling laws of observables in its proximity. In particular, we present here the calculations of the first-order corrections to the two-point functions of the theory, generalizing it to a $O(n)$-symmetric model, i.e., the one in which the order parameter $\vec{\varphi}$ is a $n$-vector whose components $\varphi_i$ satisfy Langevin equations of the form \reff{eq:Langevin}.
For the case $n=1$, the resulting predictions for the emerging scaling behaviour were in fact confirmed by Monte Carlo simulations, as briefly reported in reference~\cite{M-class}.
\change{Since we have assumed --- motivated by the central limit theorem --- the noise $\eta$ to be Gaussian, 
we can integrate over it in equation \reff{eq:act1}, thus obtaining an effective action
\begin{equation}
		S \lqq \varphi, \widetilde{\varphi} \rqq = \int_0^\infty \rd t \, \rd x_\perp  \int \rd x_{\parallel}  \, \left\{ \widetilde{\varphi} \lt \partial_t \varphi + \frac{\delta \hat{\mathcal{F}}}{\delta \varphi}  \rt - {\widetilde{\varphi}}^2   \right\}
	\label{eq:ac1}
\end{equation}
which explicitly depends  only on $\varphi$ and $\widetilde{\varphi}$. In the case at hand (see equations \reff{eq:Stree} and \reff{eq:SS}), this corresponds to
\begin{equation}
	S\left[ \varphi, \widetilde{\varphi}  \right] = \int_0^\infty \rd t \, \rd x_\perp  \int \rd x_{\parallel} \, \left\{ \widetilde{\varphi} \left[ \dot{\varphi} +  \left( r-\nabla^2 \right)\varphi +  \frac{g}{6}\varphi^2 \varphi  \right]-  \widetilde{\varphi}^2   \right\}.
		\label{eq:Taction1}
\end{equation}
for the bulk action and
\begin{equation}
	S_1 \left[ \varphi, \widetilde{\varphi}  \right] = \int_0^\infty \rd t \,   \int \rd x_{\parallel} \,  \frac{c_0}{2}\, \widetilde{\varphi} \, \varphi
		\label{eq:Taction2}
\end{equation}
for the surface one.}

\subsection{\change{Dynamical response and correlation functions}
}
Exploiting the translational invariance along the surface, one can study
the spatial Fourier transform of the two-point \emph{correlation} and \emph{response} functions
\begin{equation}
			C (\vec{k}_\parallel  ;\, x,t;\,y,s ) = \langle \varphi (\vec{k}_\parallel  ;\, x,t ) \varphi (-\vec{k}_\parallel ; \, y,s ) \rangle
\qquad \text{and} \qquad
			R (\vec{k}_\parallel  ;\,  x,t;\,y,s ) = \langle \varphi (\vec{k}_\parallel ;\, x,t ) \widetilde{\varphi} (-\vec{k}_\parallel ;\, y,s )\rangle,
			\label{eq:2point}
\end{equation}
respectively, which depend on the \change{wavevector $\vec{k}_\parallel$ (in the following denoted just by $\vec{k}$ for simplicity)} and on the distances $x$ and $y$ of the two points from the surface, in addition to the times $t$ and $s$ elapsed since the quench.
The remaining two-point function $\av{\widetilde{\varphi}\widetilde{\varphi}}$ vanishes identically \cite{Tauber2}.
For every value of $\vec{k}_\parallel$, the ordinary and special transitions correspond to the boundary conditions
$\{\varphi(x_\perp = 0, t) = 0,\widetilde{\varphi}(x_\perp = 0, t) = 0\}$ and
$\{\partial_{x_\perp} \varphi(x_\perp = 0, t) = 0,\partial_{x_\perp}\widetilde{\varphi}(x_\perp = 0, t) = 0\}$
(the latter being valid only within the Gaussian approximation) which are supplemented by the initial condition
$\varphi(x_\perp, t=0) = 0$. The resulting correlation and response function within the Gaussian approximation $[{}^{(0)}]$
turn out to be \cite{Janssen, Special, Ordinary, CritDyn-Gambassi}:
	\begin{equation}
	\begin{split}
		R^{(0)}( \vec{k};\, x,t;\,y,s ) &= R^{(0)}_{(\mathrm{b,eq})}( \vec{k};\,x-y, t-s)
\pm R^{(0)}_{(\mathrm{b,eq})}(\vec{k};\, x+y, t-s ) \\[2mm]
		&= \frac{\theta(t-s)}{\sqrt{\pi (t-s)}}   \exp\left\{-k^2\left( t-s \right)-\frac{  x^2 + y^2}{4\left(  t-s  \right)}\right\}  \,   f_{\pm} \lt \frac{xy}{2(t-s)} \rt   ,
	\label{eq:treeR}
	\end{split}
	\end{equation}
	\begin{equation}
	\begin{split}
		& C^{(0)}   ( \vec{k};\,x,t;\,y,s  ) = C^{(0)}_{(\mathrm{b,eq})} (\vec{k};\, x-y, t-s  ) - C^{(0)}_{(\mathrm{b,eq})} ( \vec{k};\,x-y, t+s  ) \\
		 & \phantom{C^{(0)}   ( \vec{k};\,x,t;\,y,s  ) =}
\pm \left[ C^{(0)}_{(\mathrm{b,eq})} (\vec{k};\, x+y, t-s ) -  C^{(0)}_{(\mathrm{b,eq})} (\vec{k};\, x+y, t+s  ) \right]  \\[2mm]
		 & \phantom{C^{(0)}   ( \vec{k};\,x,t;\,y,s  )}  =\int_{\abs{t-s}}^{t+s} \frac{\rd u}{\sqrt{\pi u}} \,\,
\exp\left\{-k^2 u -\frac{x^2 + y^2}{4u}\right\} f_{\pm} \lt \frac{xy}{2u} \rt ,
	\label{eq:treeC}
	\end{split}
	\end{equation}
where the upper and lower signs refer to the special and ordinary phase transitions, respectively,
and $f_{\pm} (\alpha) = (\rme{\alpha} \pm \rme{-\alpha})/2$; $C^{(0)}_{(\mathrm{b, eq})}$, $R^{(0)}_{(\mathrm{b, eq})}$
are the corresponding functions in the \emph{bulk} at \emph{equilibrium}, i.e., in the absence of boundaries, which are given by
		\begin{equation}
		\sysb{l}
			R^{(0)}_{(\mathrm{b,eq})}( \vec{k} ;\,  \Delta x, \Delta t) = \theta(\Delta t)
			\rme{-k^2\Delta t-( \Delta x)^2/(4\Delta t)}\big/{\sqrt{4\pi \Delta t}} ,  \\[3mm]
			C^{(0)}_{(\mathrm{b,eq})}(  \vec{k} ;\, \Delta x, \Delta t) = \int_{\left| \Delta t\right|}^{+\infty} \rd u \  R^{(0)}_{(\mathrm{b,eq})}(\Delta x, u;\,\vec{k}) .
		\syse
	\label{eq:2pointbe}
	\end{equation}
Here, $\theta(t>0)=1$ and $\theta(t\leqslant 0) =0$ ensures the causality of
 the response function.
In order to highlight the effect of the boundaries on the collective dynamics, hereafter we focus on the case in which the system is
quenched right at its critical point, i.e., we fix $r=r_\textrm{c}$ for $t>0$, where $r_\textrm{c}$ is the critical value of the parameter $r$; beyond the Gaussian approximation $r_\textrm{c}$ still vanishes if the analysis is done by using dimensional regularization to calculate the relevant integrals.
Note that  $C^{(0)}_{(\mathrm{b,eq})}$ in equation \reff{eq:2pointbe} is constructed from $R^{(0)}_{(\mathrm{b,eq})}$ via the classical \emph{fluctuation-dissipation theorem}, which holds in equilibrium \cite{FDT-rev}.

\subsection{\change{First-order corrections}}

\begin{figure}[!h]
		\centering
		\includegraphics[width=0.4\columnwidth]{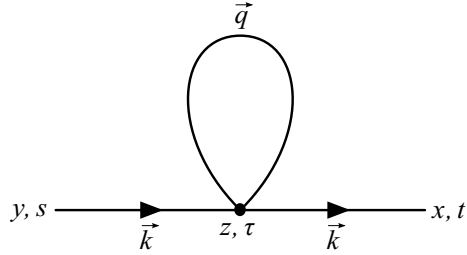}
		\caption{One-loop correction $R^{(1)}(\vec{k};x,t; y,s)$ to the response function. Undirected and directed lines correspond to $C^{(0)}$ and $R^{(0)}$, respectively; the arrows point towards later times according to the causal structure of the response.}
		\label{fig:tadpole}
		\end{figure}

The presence of the interaction $\propto g \widetilde{\varphi} \varphi^3$ in equation~\reff{eq:Taction1} can be accounted for in
perturbation theory, and gives rise to corrections to $R^{(0)}$ and $C^{(0)}$, represented by Feynman diagrams (see, e.g., reference \cite{Zinn-Justin}).
In what follows, we focus on the first non-vanishing correction  $R^{(1)}$ to the response function, since the one for the correlation function
yields the same critical exponents; this term is represented by the diagram in figure \ref{fig:tadpole},
where directed lines correspond to $R^{(0)}$ in equation \reff{eq:treeR} and undirected ones correspond to $C^{(0)}$ in \reff{eq:treeC}.
%
%
%

The corresponding expression is
\begin{equation}
		R^{(1)} ( \vec{k}; x,t; y,s ) = -\frac{n+2}{6}g  \int_0^\infty \rd z \, \int_s^t\rd \tau \,         R^{(0)} (\vec{k};\, x,t;\,z,\tau  )    \,  R^{(0)} (\vec{k};\, z,\tau ;\,y,s  ) \, B(z,\tau),
		\label{eq:defR1}
\end{equation}
where
\begin{equation}
	B\left( z,\,\tau  \right) = \int \dd{d-1}{q} \, C^{(0)} \left(\vec{q};\, z,\tau;\,z,\tau  \right)
\label{eq:bubble}
\end{equation}
is the ``bubble'' in the diagram of figure \ref{fig:tadpole}.
The different contributions coming from the four terms in equation \reff{eq:treeC}
--- labeled below by indices 0, 1, 2, 3 ---  correspond to the effects of the various boundaries.
Accordingly, $B$ decomposes as $B\left( z,\,\tau  \right) = \sum_{i=0}^3 \varepsilon_i\, \mathcal{B}_i$,
where
	\begin{equation}
	\mathcal{B}_i
	= \int \dd{d-1}{q}\,\, C^{(0)}_{(\mathrm{b,eq})}\left(\vec{q};\, Z_i, T_i  \right) = \int \dd{d-1}{q} \, \int_{T_i}^\infty \frac{\rd u}{\sqrt{4\pi u}}  \rme{-q^2 u - Z_i^2/(4u)},
	\end{equation}
with $(\varepsilon_0,Z_0,T_0) = (1,0,0)$, $(\varepsilon_1,Z_1,T_1) = (-1,0, 2\tau)$, $(\varepsilon_2,Z_2,T_2) = (\pm 1,2z,0)$, and $(\varepsilon_3,Z_3,T_3) = (\mp  1,2z,2\tau)$.
The upper and lower signs distinguish the special from the ordinary transition. Within dimensional regularisation \cite{Zinn-Justin, Diehl-rev}
one  finds $\mathcal{B}_0 =0$ and
	\begin{subequations}
	\begin{align}
	&\mathcal{B}_1
	= ( 4\pi)^{-d/2} ( 2\tau)^{1-d/2}/(d/2-1) , \label{eq:B1}  \\[1mm]
	&\mathcal{B}_2
	= ( 4\pi)^{-d/2} \,z^{2-d}\, \Gamma\left(d/2 - 1  \right),  \label{eq:B2}  \\[1mm]
	&\mathcal{B}_3
	= ( 4\pi)^{-d/2}\,z^{2-d} \, \gamma\left(d/2 - 1 , z^2/(2\tau) \right),  \label{eq:B3}
	\end{align}	
	\end{subequations}
where $\gamma\left( \alpha ,\,w \right) = \int_0^{w} \rd z\,\, z^{\alpha-1} \,\rme{-z}$ is the incomplete gamma function.
Each $\mathcal{B}_i(z,\tau)$, once integrated as indicated on the r.h.s. of equation~\reff{eq:defR1},
yields a contribution $\mathcal{R}_i$, in terms of which
\begin{equation}
	R^{(1)} = -\frac{n+2}{6}g \suml{i=1}{3} \varepsilon_i \, \mathcal{R}_i.
	\label{eq:defR}
\end{equation}
In what follows, we set for simplicity $\vec{k}=0$ and $t>s$; furthermore, we  focus on the asymptotic behaviour of the $\mathcal{R}_i$s in the proximity of the spatial ($y=0$) and temporal ($s=0$) boundaries.
Note that within the present perturbative expansion, an algebraic behaviour $\sim x^\alpha$ with power $\alpha = \alpha_0 + \alpha_1 g + \order{g^2}$
is signaled by a logarithmic term, because
	\begin{equation}
		x^{\alpha} = x^{\alpha_0 + g\alpha_1 +\order{g^2} } = x^{\alpha_0} \left[1 + g\alpha_1 \ln x + \order{g^2}\right].
	\label{eq:expexp}
	\end{equation}
	
\subsubsection{
\change{Bulk initial-slip contributions}}	
	
By using the relation $\int_0^\infty \! \rd z \, \, R^{(0)} \left( x,t;\,z,\tau  \right) R^{(0)} \left( z,\tau ;\,y,s  \right) = R^{(0)} \left( x,t;\,y,s  \right)$, valid for $ s<\tau<t$, one readily finds
\begin{equation}
	\begin{split}
		&\mathcal{R}_1 =  \int_0^\infty \! \rd z \int_s^t \! \rd \tau \,  R^{(0)} \left( x,t;\,z,\tau  \right) R^{(0)} \left( z,\tau ;\,y,s  \right) \mathcal{B}_1(\tau)
		 \\[2mm]
		&\phantom{\mathcal{R}_1 }= \frac{2 \lt 8\pi \rt^{-d/2 } }{d/2-1   } \lqq \frac{t^{2-d/2} - s^{2-d/2}   }{2 - d/2} \rqq R^{(0)}\left( x,t;\,y,s  \right)
	\end{split}
	\label{eq:Rf1}
	\end{equation}
	which, for $d=4$, yields
	\begin{equation}
		\mathcal{R}^{(1)}_1 \lt x,t;\, y,s \rt = \frac{1}{2\left( 4\pi \right)^2}\ln (t/s) \; R^{(0)}\left( x,t;\,y,s  \right).
	\label{eq:A1}
	\end{equation}
This function is singular only for $s \to 0$; thus, it identifies an initial-slip divergence, independently of the spatial properties. In particular,
according to equation \reff{eq:expexp}, it can be reinterpreted as
an algebraic law
	\begin{equation}
		R^{(1)} (\ldots , s) \sim s^{-a} R^{(0)} (\ldots , s) =  s^{-a_0} \left [ 1 - g a_1 \ln s + \order{g^2}\right] R^{(0)} (\ldots, s) \quad \mbox{for} \quad s \to 0,
	\label{eq:aseries}
	\end{equation}
where the exponent $a=a_0+a_1g + \order{g^2}$, has
$a_0=0$ and (see equation \reff{eq:defR}) $a_1 =  (n+2)/[12 (4\pi)^2]$. The leading scaling behaviour of the system is characterized by the fact that the associated coupling constants  assume their corresponding fixed-point values, which are reached under RG transformations. In the present case and for $d=4-\epsilon < 4$, $g$ reaches the Wilson-Fisher fixed point $g = g^\ast = 3 (4\pi)^2 \epsilon/(n+8) + \order{\epsilon^2 }$ \cite{Zinn-Justin, Ordinary, Special} and, correspondingly, the exponent $a$ is
	\begin{equation}
		a = \frac{n+2}{n+8} \frac{\epsilon}{4} + \order{\epsilon^2},
	\label{eq:hata1}
	\end{equation}
which agrees with previous predictions (see reference \cite{Janssen}). Thereby, this term completely encodes the
behaviour associated with the temporal boundary.

\subsubsection{
\change{Equilibrium surface effects}}

The second term $\mathcal{R}_2$ reads, instead (see equations~\reff{eq:defR1} and \reff{eq:treeR}), 	
\begin{equation}
	\begin{split}
		\lt 4\pi \rt^{d/2} \mathcal{R}_2 &=  \int_0^\infty \rd z \int_s^t \rd \tau \, \frac{z^{2-d} \, \Gamma \lt d/2 - 1 \rt}{\sqrt{\pi^2 (t- \tau) (\tau - s)}} \, f_{\pm} \lt \frac{xz}{2(t-\tau)} \rt  f_{\pm} \lt \frac{yz}{2(\tau - s)} \rt
		\\[2mm]
		&\quad\quad\quad\times \exp \left\{   {-\frac{x^2}{4(t-\tau)} - \frac{y^2}{4(\tau-s)}- \frac{z^2}{4} \frac{t-s}{(t-\tau)(\tau - s) } } \right\} .
	\label{eq:R2first}
	\end{split}
	\end{equation}
After the changes of variables $z = 2l \sqrt{(t-\tau)(\tau - s)/(t-s)}$ and $\tau = (t-s)\vartheta + s$, this expression
becomes
\begin{equation}
	\begin{split}
	 \Omega_d  \int_0^\infty \rd l \int_0^1 \rd \vartheta \, \frac{\rme{-[\tilde{x}^2/(1-\vartheta) + \tilde{y}^2/\vartheta]/4}}{\lqq \vartheta (1-\vartheta)  \rqq^{d/2-1}} l^{2-d} \rme{-l^2 }  f_{\pm} \lt \tilde{x}l \sqrt{ \frac{\vartheta}{1-\vartheta}} \rt  f_{\pm} \lt \tilde{y} l \sqrt{ \frac{1-\vartheta}{\vartheta} } \rt,
	\end{split}
	\label{eq:PR1}
	\end{equation}
with $\Omega_d = 2^{3-d} \pi^{-1} \Gamma(d/2-1) \Delta t^{(3-d)/2}$,
$\tilde{x} = x/ \sqrt{\Delta t}$, $\tilde{y} = y/ \sqrt{\Delta t}$ and $\Delta t  = t-s$.

Recalling that $d = 4 - \epsilon$ and that we are employing dimensional regularisation \cite{Gelfand-Shilov}, the integral over $l$ in the previous equation should actually be interpreted as follows:
	\begin{equation}
		\mathcal{I} \equiv \int_0^\infty \rd l \, l^{-2+\epsilon} \left[ \rme{-l^2} f_{\pm} (Al) f_{\pm} (Bl)  - f_{\pm} ^2(0) \right],
	\label{eq:G-S}
	\end{equation}
	with $A = \tilde{x} \sqrt{\vartheta/(1-\vartheta)}$ and $B = \tilde{y} \sqrt{(1-\vartheta)/\vartheta}$.

Due to the fact that the integral is regular for $\epsilon \to 0$,
we can conveniently fix $d=4$. We now use the identity $2 f_{\pm} (Al) f_{\pm} (Bl) =  \cosha{(A+B)l } \,  \pm  \,  \cosha{(A-B)l } $
and express the hyperbolic cosines as $\cosh x = \sum_{m=0}^\infty x^{2m} / (2m)!$;
by recalling that $2\int_0^\infty \rd l \,l^{2m-2} (\re^{-l^2} - \delta_{m,0}) = \Gamma(m-1/2)$
we are able to calculate the integrals over $l$ and find
	\begin{equation}
	\begin{split}
		\lt 4\pi \rt^{2} \mathcal{R}_2 & = \frac{\Omega_4 }{4} \rme{-(\tilde{x}^2 + \tilde{y}^2)/4  }
\suml{m=0}{\infty} \frac{\Gamma \lt m-\frac{1}{2} \rt}{(2m)!}
\int_0^1 \frac{\rd \vartheta}{\vartheta (1-\vartheta)}
\exp  \left\{   -\frac{\tilde{x}^2}{4} \frac{\vartheta}{1-\vartheta}   - \frac{\tilde{y}^2}{4} \frac{1-\vartheta}{\vartheta} \right\}
		\\[2mm]
		& \quad\quad\quad\times\lqq \lt \tilde{x} \sqrt{\frac{\vartheta}{1-\vartheta}} + \tilde{y} \sqrt{\frac{1-\vartheta}{\vartheta}} \rt^{2m} \pm  \lt \tilde{x} \sqrt{\frac{\vartheta}{1-\vartheta}} - \tilde{y} \sqrt{\frac{1-\vartheta}{\vartheta}} \rt^{2m}  \rqq.
	\end{split}
	\label{eq:betam1}
	\end{equation}
With an additional change of variable $\sqrt{(1-\vartheta)/\vartheta} = \sqrt{\tilde{x}/\tilde{y}} \, \beta$,
the integral over $\vartheta$ above becomes 	
\begin{equation}
		 2\lt  \tilde{x} \tilde{y} \rt^m \int_0^\infty \rd \beta\; \beta^{-1}
		 \re^{-\tilde{x} \tilde{y}(\beta^2 + \beta^{-2})/4}
		 \lqq \lt \beta + \beta^{-1} \rt^{2m} \pm  \lt \beta - \beta^{-1} \rt^{2m}  \rqq.
\label{eq:beta}
\end{equation}
It can be proved that the resulting series in equation \reff{eq:betam1} is pointwise convergent for $\tilde{x} \tilde{y} > 0$ and that the only
singular term for $ \tilde{x} \tilde{y} = 0$ is the one with $m=0$, i.e.,
	\begin{equation}
		2(1\pm1)\EGamma{-1/2} \int_0^\infty \rd \beta \, \beta^{-1} \rme{-\tilde{x} \tilde{y} (\beta^2 + \beta^{-2})/4 }
		= -4 (1\pm 1)\sqrt{\pi} \BK{0}{\tilde{x} \tilde{y}/2},
	\label{eq:term21}
	\end{equation}
where $K_\nu$ is the modified Bessel function of the second kind, whose asymptotic behaviour
is $K_0 (z\to 0) \sim -\ln z$.
For the special transition, this implies that, in the limit $\tilde{x} \tilde{y} \to 0$, equation \reff{eq:beta} is dominated by the logarithmic singularity
	\begin{equation}
	\begin{split}
		\lt 4\pi \rt^{2} \mathcal{R}^{(1)}_2 \big|_{{\rm div}}  \sim 2\sqrt{\pi} \, \Omega_4  \, \rme{-(\tilde{x}^2 + \tilde{y}^2 )/4} \lna{ \tilde{x} \tilde{y} } = \frac{2}{\sqrt{4\pi \Delta t }} \rme{-(\tilde{x}^2 + \tilde{y}^2 )/4} \lna{ \tilde{x} \tilde{y} },
	\end{split}
	\label{eq:beta1}
	\end{equation}
which, after an allowed multiplication by $\cosha{\tilde{x} \tilde{y}/2} \approx 1$, can be cast in the form
	\begin{equation}
		\mathcal{R}^{(1)}_2 \big|_{{\rm div.}}  \sim \frac{1}{\lt 4\pi \rt^2} \lna{\frac{xy}{\Delta t}} R^{(0)} \lt x,t;y,s \rt.
	\label{eq:R22}
	\end{equation}
As in the  case of $\mathcal{R}_1^{(1)}$ in equation~\reff{eq:A1}, this behaviour can be traced back to
the emergence of an algebraic law
	\begin{equation}
		R^{(1)} (\ldots ) \sim \lt \frac{xy}{\Delta t} \rt^{b} R^{(0)} (\ldots) =  \lt \frac{xy}{\Delta t} \rt^{b_0} \lqq 1 + g b_1 \lna{\frac{xy}{\Delta t}}  + O(g^2)\rqq   R^{(0)} (\ldots) \ \mbox{for}\   xy \to 0,
	\label{eq:bseries}
	\end{equation}
with exponent $b = b_0 + b_1 g + \order{g^2}$;
for the \emph{special} transition, $b^{(\mathrm{sp})}_0 = 0$,  $b^{(\mathrm{sp})}_1 = -(n+2)/[6(4\pi)^2]$, i.e.,
\begin{equation}
b^{(\mathrm{sp})}  = -\frac{n+2}{n+8} \frac{\epsilon}{2} + \order{\epsilon^2}
\label{eq:bsp}
	\end{equation}
at the Wilson-Fisher fixed point; this algebraic behaviour reproduces the one predicted
for the surface scaling \cite{Diehl-rev, Special}.
Accordingly, this term correctly and completely captures the surface divergence in the special case.

As mentioned in section~\ref{sec:concl}
(see, e.g., equation \reff{eq:comp-exp}), $b$ can be actually expressed \cite{Diehl-rev} in terms of the bulk and surface exponents $\beta$ and $\beta_1$, respectively, introduced in equation \reff{eq:diffsc} for the magnetization $m$.
%
For the \emph{ordinary} transition, the two-point function obeys Dirichlet boundary conditions, which yields $b_0^{(\mathrm{ord})}
= 1$ while the first (i.e., $m=0$) term of the series in equation \reff{eq:beta} identically vanishes
and the leading contribution is of the order $\tilde{x} \tilde{y} \ln (\tilde{x} \tilde{y})$, coming entirely from the $m=1$ term.
In this case, an analogous calculation yields $b_1^{(\mathrm{ord})}  = -(n+2)/[6 (4\pi)^2]$ and
	\begin{equation}
		b^{(\mathrm{ord})} = 1 -\frac{n+2}{n+8} \frac{\epsilon}{2} + \order{\epsilon^2},
		\label{eq:bord}
	\end{equation}
	which correctly reproduces the previously-known results for the ordinary transition \cite{Ordinary}.
	

\subsubsection{\change{New 
singularities and edge corrections to scaling}}
Finally, we consider the third term $\mathcal{R}_3^{(1)}$; its expression is
the same as in equation \reff{eq:R2first} with  $\EGamma{d/2-1}$ replaced by the incomplete gamma function $\gamma(d/2 - 1, z^2/(2\tau))$; by introducing the same change of variables as in equation \reff{eq:R2first}, one arrives at	
	\begin{equation}
	\begin{split}
		\lt 4\pi \rt^{d/2} \mathcal{R}^{(1)}_3 & = \frac{\Omega_d }{\EGamma{d/2-1}} \int_0^\infty \rd l \int_0^1 \rd \vartheta  \,    \exp    \left\{-\frac{1}{4} \lt \frac{\tilde{x}^2}{1-\vartheta} + \frac{\tilde{y}^2}{\vartheta} \rt  \right\}  \,  \lqq \vartheta (1-\vartheta)  \rqq^{1-\frac{d}{2}}  \,  l^{2-d} \, \rme{-l^2 }  \\[2mm]
		&\quad\quad\quad\quad \times    f_{\pm} \lt \tilde{x}l \sqrt{ \frac{\vartheta}{1-\vartheta}} \rt  f_{\pm} \lt \tilde{y} l \sqrt{ \frac{1-\vartheta}{\vartheta} } \rt \,  \gamma \lt \frac{d}{2}-1, \frac{2l^2\vartheta(1-\vartheta)}{\vartheta + s/\Delta t} \rt  .
	\end{split}
	\label{eq:RR3}
	\end{equation}
Since $\gamma(\alpha,w\to 0)$
vanishes as $\sim w^{\alpha}$, it constitutes a sufficient regularisation to make the integral in $l$ convergent.
One can, therefore, set $d=4$ from the outset, noticing that, correspondingly, $\gamma (1,w) = 1- \rme{-w}$. Thus,
this integral becomes
	\begin{equation}
		\wt{\mathcal{I}} =    \int_0^\infty \rd l \, \rme{-l^2} \, f_\pm (Al) \, f_\pm (Bl) \, \frac{1-\rme{-Cl^2}}{l^2},
	\end{equation}
with the same $A$ and $B$ as in equation \reff{eq:G-S}, $C = 2\vartheta  (1-\vartheta) /(\vartheta + \tilde{s}) $ and $\tilde{s} = s/\Delta t$.
In order to extract the possible logarithmic contributions localised at the boundaries $x = 0$, $ y = 0$, $s = 0$, we
rewrite $\wt{\mathcal{I}}$ as follows:
	\begin{equation}
	\begin{split}
		\wt{\mathcal{I}}_1 +  \wt{\mathcal{I}}_2 = \int_0^\infty \rd l \, \rme{-l^2} \left[ f_\pm (Al) \, f_\pm (Bl)  - f^2_\pm(0) \right] \frac{1-\rme{-Cl^2}}{l^2}   \,  +   \int_0^\infty \rd l \, \rme{-l^2} f^2_\pm(0) \frac{1-\rme{-Cl^2}}{l^2}
	\end{split}
	\label{eq:I1I2}
	\end{equation}
and note that $\wt{\mathcal{I}}_1$ represents a more regular version of $\mathcal{I}$ (see equation \reff{eq:G-S});
thereby, for the special ($+$) case, this term is always regular and one can entirely focus on $\wt{\mathcal{I}}_2$. On the other hand, in the ordinary case $f_{-} (0) = 0$ and $\wt{\mathcal{I}}_2$ vanishes.
Again, by applying to $\wt{\mathcal{I}}_1$ what we have found above for $\mathcal{I}$, one can restrict to considering the first order in the
expansion of the hyperbolic functions, i.e., $f_-(Al) f_-(Bl) = \sinha{Al} \sinha{Bl} \simeq  AB l^2$.
Remarkably, the analysis reported below for the special case can be similarly repeated for the ordinary transition.
Making use of the identity
\begin{equation}
		\int_0^\infty \rd l \, \rme{-l^2} (1-\rme{-Cl^2})/l^2 = \sqrt{\pi} \left(\sqrt{1+C} -1\right)
\end{equation}
and introducing $\mu = \sqrt{(1-\vartheta)/\vartheta} $ one finds
	\begin{equation}
	\begin{split}
		\lt 4\pi \rt^{2} & \mathcal{R}^{(1)}_3 \big|_{{\rm div.}}  = -2\sqrt{\pi}  \,\,  \Omega_4  \, \, \rme{-(\tilde{x}^2 + \tilde{y}^2)/4} \int_0^\infty \frac{\rd \mu}{\mu} \,  \exp  \left\{-\frac{1}{4} \lt \frac{\tx^2}{\mu^2} + \ty^2 \mu^2 \rt   \right\} \times \\[2mm]
		&\times \lqq 1- \sqrt{1+\frac{2\mu^2}{(\mu^2 + 1) \lt 1 + \tilde{s} (\mu^2 + 1) \rt}}  \rqq  \equiv -2\sqrt{\pi}  \,\,  \Omega_4  \, \, \rme{-(\tilde{x}^2 + \tilde{y}^2)/4} \widetilde{Q}(\tilde{x}, \tilde{y},\tilde{s}).
	\end{split}
	\label{eq:divR3}
	\end{equation}
The argument of the square brackets in the expression above behaves as $-\mu^2$ for $\mu \to 0$, whereas for $\mu \to \infty$ it vanishes as $\mu^{-2} /\tilde{s} $ for $s>0$ and approaches $1-\sqrt{3}$ for $s=0$. Therefore, even in the absence of the exponential (i.e., for $x=y=0$) the integral is convergent for every $s>0$. We also notice that the integral is still finite for $s=x=0$, $y>0$, as the exponential regularises the behaviour at $\mu \to \infty$. Thus, a singularity can be obtained only for $y=s=0$, independently of $x$. Because of this, we introduce the \change{representation
	\begin{equation}
		\tilde{y}^2 = u \cos \alpha \quad \mbox{and} \quad \tilde{s} = u \sin \alpha,
		\label{eq:radial}
	\end{equation}
in terms of $u= \sqrt{\tilde{s}^2 + (\tilde{y}^2)^2}$, which acts as an effective ``radial'' coordinate in the space of directions orthogonal to the spatial and temporal boundaries, and the corresponding ``polar angle'' $\alpha$. Note that we have also} accounted for the different scaling of time and space, described by the dynamical exponent $z = 2 + O(\epsilon^2)$ \cite{Janssen}. Accordingly, one has $\widetilde{Q} (\tilde{x},\sqrt{u \cos \alpha} , u \sin \alpha ) \equiv Q(u, \alpha)$, where we do not explicitly indicate the dependence on $\tilde{x}$ since it plays no significant role.

As in the previous cases, we expect a logarithmic behaviour $Q(u,\alpha) = f(\alpha) \ln u + \order{u^0}$ to emerge for $u \to 0$.
In order to highlight it and calculate the coefficient $f(\alpha)$, we derive this function with respect to $u$ and introduce
$\mu = \gamma/\sqrt{u}$, which yields $Q'(u,\alpha) \equiv \partial_u Q(u, \alpha) = [J_1 (u,\alpha) + J_2(u,\alpha)]/u$ with
	\begin{subequations}
	\begin{align}
		J_1 (u,\alpha) & = -\frac{\cos\alpha}{4} \int_0^\infty  \rd \gamma \, \,  \gamma \,\,  \rme{-\frac{1}{4} \lt u \tx^2/\gamma^2 + \gamma^2 \cos\alpha   \rt }    \lqq  1 - \sqrt{1 + \frac{2\gamma^2}{(\gamma^2 + u) \lqq 1 + (\gamma^2 + u) \sin\alpha  \rqq}}  \rqq, \\[2mm]
		J_2 (u,\alpha) & = \sin\alpha  \int_0^\infty  \rd \gamma \, \,  \gamma \,\,  \frac{\rme{-\frac{1}{4} \lt u \tx^2/\gamma^2 + \gamma^2 \cos\alpha   \rt }}{[1 + (\gamma^2 + u)  \sin\alpha]^2}  \lqq 1 + \frac{2\gamma^2}{(\gamma^2 + u) \lqq 1 +  (\gamma^2 + u)  \sin\alpha\rqq}  \rqq^{-1/2}.
	\end{align}
	\end{subequations}
In these terms, $f(\alpha)$ is given by $J_1 (0,\alpha) + J_2 (0,\alpha)$, provided it is finite.
The first addend $J_1(0,\alpha)$ can be rewritten as follows:
\begin{equation}
	\begin{split}
		J_1 (0,\alpha) = \frac{1}{2} \int_0^\infty \rd \gamma \,  \lt  \partial_\gamma \,   \rme{-\frac{1}{4} \gamma^2 \cos \alpha}  \rt \lqq  1 - \sqrt{1+ \frac{2}{ 1 +  \gamma^2  \sin\alpha}}       \,\rqq = -\frac{1}{2} + \mathcal{J} (\alpha),
	\end{split}
	\end{equation}
where $\mathcal{J} (\alpha)$ denotes the contribution of the second term within the brackets (i.e., the square root).
We then rewrite $J_2(0,\alpha)$ as follows:
	\begin{equation}
	\begin{split}
		&  J_2(0, \alpha) =
-\int_0^\infty \rd \gamma  \, \gamma \,\,   \rme{-\frac{1}{4} \gamma^2 \cos\alpha} \frac{1}{2\gamma} \partial_\gamma \lt   1 + \frac{2}{1+ \gamma^2 \sin\alpha}   \rt^{\frac{1}{2}}   =   \frac{\sqrt{3}}{2}  - \mathcal{J} (\alpha),
	\end{split}
	\end{equation}
where the last equality follows from an integration by parts.
This confirms that the divergence of $Q(u,\alpha)$ for $u \to 0$ is indeed logarithmic in nature.
Moreover, it proves that the coefficient $f(\alpha) = (  \sqrt{3}-1)/2$
is actually independent of the choice of $\alpha$, which means that the divergence
is the same when approaching the edge from any ``direction'' in the $y^z-s$ plane.
Thus, the divergent part \reff{eq:divR3} can be rewritten as follows:
	\begin{equation}
		\lt 4\pi \rt^{2} \mathcal{R}^{(1)}_3 \big|_{{\rm div.}} \sim     -  \frac{1}{\sqrt{4\pi\Delta t}}  \rme{-\tx^2/4}  \frac{\sqrt{3}-1}{2}   \ln u  = -  \frac{\sqrt{3}-1}{2}  \ln u \,\, R^{(0)} (x,t;0,0),
	\label{eq:thirdlast}
	\end{equation}
which entails the emergence of a novel, algebraic law $R(x,t, u) \sim  u^{-\theta_E}$ in the vicinity of the edge (i.e., for $u \to 0$) with $\theta_E = \theta_{E,0} + \theta_{E,1} g + \order{g^2}$. The actual exponent can be inferred by multiplying the expression above by $ (n+2) g/6$
which yields $\theta_{E,0}^{(\mathrm{sp})}=0$ and $\theta_{E,1}^{(\mathrm{sp})}= (4 \pi)^2 g (n+2)  (\sqrt{3} -1)/12$, and, therefore,
\begin{equation}
		\theta_E^{(\mathrm{sp})} =
		\frac{n+2}{n+8} \lt \frac{\sqrt{3}-1}{4}  \rt \epsilon + \order{\epsilon^2}
		\label{eq:thEsp}
\end{equation}
at the Wilson-Fisher fixed point.
As stated above, for the ordinary transition one can focus on the first order in the expansion of $f_-$, which leads to an expression similar to equation \reff{eq:divR3} with the corresponding function $Q$ given by (to be compared with equation \reff{eq:divR3})
\begin{equation}
	\int_0^\infty \frac{\rd \mu}{\mu} \,  \exp  \left\{-\frac{1}{4} \lt \frac{\tx^2}{\mu^2} + \ty^2 \mu^2 \rt   \right\}   \, \lqq 1-         \lt   1+\frac{2\mu^2}{(\mu^2 + 1) \lt 1 + \tilde{s} (\mu^2 + 1) \rt} \rt^{-1/2}  \rqq.
\end{equation}
By repeating the analysis outlined above, one eventually finds
\begin{equation}
		\theta_E^{(\mathrm{ord})} =
		 \frac{n+2}{n+8}   \lt \frac{1}{\sqrt{3}}  -1  \rt \frac{\epsilon}{4} + \order{\epsilon^2},
		\label{eq:thEord}
\end{equation}	
which, contrary to $\theta_E^{(\mathrm{sp})}$, turns out to be negative.

\section{\change{Scaling forms and comparison with previous results}} 

\change{The calculations above identify a triplet of (field-theoretical) singularities localised at the initial time, at the surface and at the edge, respectively; this implies that the two-point functions are bound to display different power-law behaviours in the proximity of each boundary. Accounting for this fact, the most general scaling forms one can write for the two-point expectations \reff{eq:2point} are as follows:
		\begin{subequations}
		\begin{align}
			C(x,t;y,s) = & \Delta t^{1-\eta}    \left( \frac{s}{t} \right)^{1-\theta} \left( \frac{{\cal{A}}^2 xy}{\Delta t^{2/z}}  \right)^{(\beta_1 - \beta)/\nu}  \left[\frac{({\cal{A}}y)^z + s}{\Delta t}  \right]^{-\theta_E}   F_C \lt \frac{({\cal{A}}x)^z}{\Delta t},\frac{({\cal{A}}y)^z}{\Delta t}, \frac{s}{t} \rt,
			\label{eq:Csc} \\[3mm]
			R( x,t;\,y,s) = &   \Delta t^{-\eta}       \left( \frac{s}{t}  \right)^{-\theta} \left(\frac{{\cal A}^2 xy}{\Delta t^{2/z}}  \right)^{(\beta_1 - \beta)/\nu} \left[ \frac{({\cal A}y)^z+s}{ \Delta t }  \right]^{-\theta_E} F_R \lt \frac{({\cal A}x)^z}{\Delta t},\frac{({\cal A}y)^z}{\Delta t}, \frac{s}{t} \rt    ,
			\label{eq:Rsc}
		\end{align}
		\end{subequations}
(with $\Delta t = t-s$) for the correlation and response functions, respectively, which are valid for $t>s$ and where the dependence on the wavevector $\vec{k}_\parallel$ parallel to the surface is here understood in the form of an additional dependence on a scaling variable ${\cal A}^{-1}k_\parallel \Delta t^{1/z}$.
In the expressions above, $\cal{A}$ represents a non-universal constant which
accounts for the difference in physical dimension between space and time and is required to construct a meaningful "radial" coordinate, whilst $F_{R/C}$ are scaling functions which remain finite for vanishing arguments (i.e., $0< \abs{F_{R/C}(0,0,0)} < \infty$). These functions are \emph{universal} up to an overall multiplicative normalization constant which can be suitably fixed, for example, by comparing with their corresponding expression in the spatial bulk \cite{CritDyn-Gambassi}.
The multiplicative factors appearing in front of $F_{R/C}$
capture, from left to right, the scaling dimensions of the two-point functions, the initial slip behaviour, the surface behaviour and the corrections due to the edge. The corresponding exponents, which are here expressed with the notation most commonly adopted in the literature,
are related to the ones in equations \reff{eq:hata1}, \reff{eq:bsp} and \reff{eq:bord} via
\begin{equation}
	\theta = a   \quad \text{and} \quad    (\beta^{(\mathrm{ord}/\mathrm{sp})}_1 - \beta)/\nu = b^{(\mathrm{ord}/\mathrm{sp})},
	\label{eq:comp-exp}
\end{equation}
while $\theta_E$ is reported in equations \reff{eq:thEsp} and \reff{eq:thEord}.}

\change{In references \cite{Ritschel, Sengupta, CritDyn-Gambassi}, ansatzes for the scaling forms have been formulated on the basis of a power-counting argument which actually rules out the presence of non-trivial edge contributions. In our notation,
they read
		\begin{subequations}
		\begin{align}
			C(x,t;y,s) = & \Delta t^{1-\eta}    \left( \frac{s}{t} \right)^{1-\theta} \left( \frac{{\cal{A}}^2 xy}{\Delta t^{2/z}}  \right)^{(\beta_1 - \beta)/\nu}   f_C \lt \frac{({\cal{A}}x)^z}{\Delta t},\frac{({\cal{A}}y)^z}{\Delta t}, \frac{s}{t} \rt,
			\label{eq:Ccsc} \\[3mm]
			R( x,t;\,y,s) = &   \Delta t^{-\eta}       \left( \frac{s}{t}  \right)^{-\theta} \left(\frac{{\cal A}^2 xy}{\Delta t^{2/z}}  \right)^{(\beta_1 - \beta)/\nu}  f_R \lt \frac{({\cal A}x)^z}{\Delta t},\frac{({\cal A}y)^z}{\Delta t}, \frac{s}{t} \rt    ,
			\label{eq:Rcsc}
		\end{align}
		\end{subequations} 	
with the requirement that $0< \abs{f_{R/C}(0,0,0)} < \infty$, which does not allow one to relate equations \reff{eq:Ccsc} and \reff{eq:Rcsc} to \reff{eq:Csc} and \reff{eq:Rsc} by simply absorbing the additional factors into $F_{R/C}$.
However, 
the behaviour at the temporal and spatial boundaries can be equally extracted from both: in fact, upon varying the time $s$ for $s \ll t, x^z , y^z$ one finds that the ``radial'' contribution becomes approximately constant 
(since $s \ll ({\cal{A}}y)^z $), hence $R(s) \sim s^{-\theta}$ and $C(s) \sim s^{1-\theta}$. Analogously, as functions of the distance $y$ for $y \ll x, t^{1/z}, s^{1/z}$, the ``radial'' coordinate becomes $\approx s/\Delta t$ and all the expressions above yield $R(y) \sim y^{(\beta_1 - \beta)/\nu}$ and $C(y) \sim y^{(\beta_1 - \beta)/\nu}$. It is, therefore, clear that the only regime in which these two predictions can be actually distinguished is the vicinity of the edge. For example, by varying $s$ in the regime $y^z \ll s \ll t$, one would find
\begin{equation}
	R(s) \sim s^{-\theta - \theta_E}  \quad \text{and} \quad C(s) \sim s^{1-\theta - \theta_E}
\label{eq:scal-noi}
\end{equation}
from equations \reff{eq:Csc} and \reff{eq:Rsc}, 
whilst
\begin{equation}
	R(s) \sim s^{-\theta } \quad \text{and} \quad C(s) \sim s^{1-\theta }
\label{eq:scal-altri}
\end{equation}
from equations \reff{eq:Ccsc} and \reff{eq:Rcsc}.
Consequently, the latter are capable of correctly capturing only the regimes in which a boundary is approached while keeping far from the other one. Since they account for all the boundaries identifiable in this system, equations \reff{eq:Csc} and \reff{eq:Rsc} should provide a complete characterization of universal features in these quantities. As we mention in the Conclusions below, numerical simulations \cite{M-class}
support the validity of equation \reff{eq:scal-noi}.}

\section{Conclusions}
\label{sec:concl}

We investigated the universal properties of a Landau-Ginzburg model with $n$-component vector order parameter, $O(n)$ symmetry, and a purely dissipative dynamics (model A) \cite{HH} in which the translational symmetries in space and time are broken by the presence of a surface and by suddenly quenching the temperature, respectively. In particular, we considered a quench from the high-temperature disordered phase to the critical point.
Previous studies of this issue within the same framework were based on the assumption --- supported by a power-counting argument --- that the intersection of these boundaries, i.e., the edge, would not introduce any novel feature in the picture \cite{Ritschel, Sengupta, CritDyn-Gambassi}.
On the other hand, the emergence of different scaling behaviours at the surface and in the bulk, both for single boundaries \cite{Diehl-rev, Ordinary, Special, Janssen} and spatial wedges \cite{Cardy-edge}, had been understood via the introduction of a new boundary field which characterizes the properties of the order parameter at the surface but has in general a different scaling dimension (see, e.g., equation \reff{eq:diffsc}), which affects the asymptotic behaviour of the observables upon approaching such an edge. In more than one spatial dimension ($d>1$), the edge constitutes a bona fide extended boundary, hence there is no apparent reason why the same logic would not apply.
The explicit calculation we have reported here in some detail proves indeed that this is the correct way of approaching the problem; as expected on these grounds, we have in fact identified a new logarithmic singularity in our perturbative expansion which is strictly localised at the edge and, therefore, can only modify the scaling behaviour of the observables in its proximity (see, e.g., equation \reff{eq:thirdlast} with the definitions \reff{eq:radial}).

Further confirmation of the presence of edge effects has been sought in reference \cite{M-class}, which reports a numerical Monte Carlo study of the classical Ising  model in three spatial dimensions (i.e., $n=1$ and $d=3$) with Glauber dynamics.
The comparison between the theoretical predictions presented here and these numerical data requires some care:
in fact, the predictions for $\theta_E$ reported here are limited to the first-order corrections
in a dimensional expansion, which are generally known not to be quantitatively accurate. However,
their qualitative features are usually more robust: for example, their sign typically dictates
the one of the entire expansion they belong to.
In the present case, it is, therefore, convenient to design the simulations in such a way as to probe the correlation and response functions in a time regime in which $s$ is varied within the window $t \gg s \gg y^z$ because this analysis would highlight an algebraic decay
 $C(\ldots, s) \sim s^{1-\theta}$ (see equation~\reff{eq:scal-altri}) in the absence of edge effects, or $C(\ldots, s) \sim s^{1-\theta - \theta_E}$ (see equation~\reff{eq:scal-noi}) in their presence. \change{(In passing we mention that $\mathcal{A}$ turns out to be of the order 1 in the numerical simulations of reference~\cite{M-class}.)}
Noticing that for the ordinary and special cases, the exponent $\theta_E$ in equations \reff{eq:thEsp} and \reff{eq:thEord} takes opposite signs, one can look for differences between the two transitions, which are expected to display faster/slower power laws with respect to the regime $s \ll y^z \ll t$, which is dominated instead by the initial-slip physics and invariably leads to observing $C(\ldots, s) \sim s^{1-\theta}$. The results of reference \cite{M-class} are in agreement with such predictions and, therefore, constitute evidence of the existence of the edge modifications to the scaling laws discussed above. As a further confirmation, a crossover from the former to the latter regime is also observed upon moving away from the surface and eventually leads to a collapse of the data on a master curve which is indeed independent of the surface transition undergone by the model.

\vspace{-7mm}

\ukrainianpart

 \title{Критична релаксація і спільний вплив просторових і часових границь}

 \author{M. Маркуцці\refaddr{label1,label2}, A. Гамбассі\refaddr{label1}}
 \addresses{
 \addr{label1} Міжнародна школа перспективних досліджень,  Трієст, Італія
 \addr{label2} Школа фізики і астрономії, Університет м. Ноттінгем,  Ноттінгем, Великобританія
 }

 \makeukrtitle

 \begin{abstract}
 \tolerance=3000%
 Ми знову розглядаємо проблему колективної нерівноважної динаміки класичної статистичної системи в критичній точці і в
 присутності поверхонь.
 Вплив порушення порізно просторової і часової трансляційної інваріантності є добре зрозумілим, тому тут ми
 зосереджуємо увагу на виникненні нетривіальної взаємодії між ними. Для цієї мети ми розглядаємо напівбезмежну
 модель з $O(n)$-симетрією і цілковито дисипативну динаміку, підготовану в невпорядкованому стані, і потім раптово заморожену
 до своєї критичної температури. Ми
 визначаємо поведінку її функції відгуку на коротких відстанях в межах теорії збурень, не спираючись на жодне припущення
 щодо форми цієї величини.
 \keywords стохастична динаміка, теорія поля з границями

 \end{abstract}

 \end{document}